\documentclass[journal,comsoc]{IEEEtran}
\makeatletter\if@twocolumn\PassOptionsToPackage{switch}{lineno}\else\fi\makeatother

\usepackage{graphicx}
\usepackage[T1]{fontenc}
\usepackage{cite}

\ifCLASSINFOpdf
\else
\fi
\usepackage{amsmath}
\usepackage[cmintegrals]{newtxmath}
\usepackage{url,multirow,morefloats,floatflt,cancel,tfrupee}
\makeatletter

\AtBeginDocument{\@ifpackageloaded{textcomp}{}{\usepackage{textcomp}}}
\makeatother
\usepackage{colortbl}
\usepackage{xcolor}
\usepackage{pifont}
\usepackage[nointegrals]{wasysym}
\makeatletter

\def\mcWidth#1{\csname TY@F#1\endcsname+\tabcolsep}

\def\cAlignHack{\rightskip\@flushglue\leftskip\@flushglue\parindent\z@\parfillskip\z@skip}
\def\rAlignHack{\rightskip\z@skip\leftskip\@flushglue \parindent\z@\parfillskip\z@skip}

\usepackage{ifxetex}
\ifxetex\else\if@twocolumn\usepackage{dblfloatfix}\fi\fi

\AtBeginDocument{
\expandafter\ifx\csname eqalign\endcsname\relax
\def\eqalign#1{\null\vcenter{\def\\{\cr}\openup\jot\m@th
  \ialign{\strut$\displaystyle{##}$\hfil&$\displaystyle{{}##}$\hfil
      \crcr#1\crcr}}\,}
\fi
}

\AtBeginDocument{%
  \@ifpackageloaded{endfloat}%
   {\renewcommand\efloat@iwrite[1]{\immediate\expandafter\protected@write\csname efloat@post#1\endcsname{}}}{\newif\ifefloat@tables}%
}%

\def\BreakURLText#1{\@tfor\brk@tempa:=#1\do{\brk@tempa\hskip0pt}}
\let\lt=<
\let\gt=>
\def\processVert{\ifmmode|\else\textbar\fi}

\@ifundefined{subparagraph}{
\def\subparagraph{\@startsection{paragraph}{5}{2\parindent}{0ex plus 0.1ex minus 0.1ex}%
{0ex}{\normalfont\small\itshape}}%
}{}

\newcommand\role[1]{\unskip}
\newcommand\aucollab[1]{\unskip}
  
\@ifundefined{tsGraphicsScaleX}{\gdef\tsGraphicsScaleX{1}}{}
\@ifundefined{tsGraphicsScaleY}{\gdef\tsGraphicsScaleY{.9}}{}
\def\checkGraphicsWidth{\ifdim\Gin@nat@width>\linewidth
	\tsGraphicsScaleX\linewidth\else\Gin@nat@width\fi}

\def\checkGraphicsHeight{\ifdim\Gin@nat@height>.9\textheight
	\tsGraphicsScaleY\textheight\else\Gin@nat@height\fi}

\def\fixFloatSize#1{}
\let\ts@includegraphics\includegraphics

\def\inlinegraphic[#1]#2{{\edef\@tempa{#1}\edef\baseline@shift{\ifx\@tempa\@empty0\else#1\fi}\edef\tempZ{\the\numexpr(\numexpr(\baseline@shift*\f@size/100))}\protect\raisebox{\tempZ pt}{\ts@includegraphics{#2}}}}

\AtBeginDocument{\def\includegraphics{\@ifnextchar[{\ts@includegraphics}{\ts@includegraphics[width=\checkGraphicsWidth,height=\checkGraphicsHeight,keepaspectratio]}}}

\DeclareMathAlphabet{\mathpzc}{OT1}{pzc}{m}{it}

\def\URL#1#2{\@ifundefined{href}{#2}{\href{#1}{#2}}}

\def\UrlOrds{\do\*\do\-\do\~\do\'\do\"\do\-}%
\g@addto@macro{\UrlBreaks}{\UrlOrds}

\edef\fntEncoding{\f@encoding}

\makeatother

\newif\ifmultipleabstract\multipleabstractfalse%
\makeatletter
\AtBeginDocument{\@ifpackageloaded{longtable}{%
\def\LT@makecaption#1#2#3{%
  \LT@mcol\LT@cols c{\hbox to\z@{\hss\parbox[t]\LTcapwidth{%
    \sbox\@tempboxa{#1{#2: } #3}%
    \ifdim\wd\@tempboxa>\hsize
      #1{#2: }\textsc{#3}%
    \else
      \hbox to\hsize{\hfil\box\@tempboxa\hfil}%
    \fi
    \endgraf\vskip\baselineskip}%
  \hss}}}
}{}}
\makeatother

\begin{document}

%

\title{Geometrical jitter and bolometric regime in photon detection by straight superconducting nanowire }
      
\author{Artem~Kuzmin,
        Steffen~Doerner,
        Mariia~Sidorova,
        Stefan~Wuensch,
        Konstantin~Ilin,
        Michael~Siegel, and 
        Alexey~Semenov\thanks{Artem~Kuzmin, Steffen~Doerner, Stefan~Wuensch, Konstantin~Ilin, Michael~Siegel are with Institut f{\"{u}}r Mikro- und Nanoelektronische Systeme (IMS) , Karlsruhe Institute of Technology, Karlsruhe, 76187, Germany (e-mail: artem.kuzmin@kit.edu). (Corresponding author: Artem~Kuzmin)}\thanks{Mariia~Sidorova, Alexey~Semenov are with Institute of Optical Sensor Systems , German Aerospace Center (DLR), Berlin, Germany}}

\maketitle 

\begin{abstract}
 We present a direct observation of the geometrical jitter in single photon detection by a straight superconducting nanowire. Differential measurement technique was applied to the 180-\ensuremath{\mu }m long nanowire  similar to those commonly used in the technology of superconducting nanowire single photon detectors (SNSPD). A non-gaussian geometrical jitter appears as a wide almost uniform probability distribution (histogram) of the delay time (latency) of the nanowire response to detected photon. White electrical noise of the readout electronics causes broadened, Gaussian shaped edges of the histogram. Subtracting noise contribution, we found for the geometrical jitter a standard deviation of \ensuremath{\approx }8.5~ps and the full width at half maximum (FWHM) of the distribution of \ensuremath{\approx }29 ps. FWHM corresponds to the propagation speed of the electrical signal along the nanowire of \ensuremath{\approx }6.2\ensuremath{\times}10\ensuremath{^{6}}~m/s or 0.02 of the speed of light.  Alternatively the propagation speed was estimated from the central frequency of the measured first order self-resonance of the nanowire. Both values agree well with each other and with previously reported values. As the intensity of the incident photon flux increases, the wide probability distribution  collapses into a much narrower Gaussian distribution with a standard deviation dominated by the noise of electronics. We associate the collapse of the histogram with the transition from the discrete, single photon detection to the uniform bolometric regime.

\end{abstract}
    

%
\IEEEpeerreviewmaketitle

\section{Introduction}
\IEEEPARstart{T}{ming} jitter is an important metric of any photon detector. The detection process itself sets its ultimate value. For detector systems with SNSPDs, experimentally measured system jitter contains inevitably contributions of electrical noise, optics, recording instruments and the detector geometry. According to the formalism developed in \cite{sidorova2017physical, sidorova2018timing}, a photon-count event is a sequence of elementary events with their particular delay times.  System jitter reveals randomness of the total delay time between the appearance of the photon at the instrument input and the emergence of the photon count. Statistics of the random delay time is described by its probability density function (PDF). Standard deviation (STD) or FWHM associated with the PDF are both the measure of the jitter. 
\\Timing jitter in superconducting nanowires has been actively investigated in a few past years both theoretically \cite{zhao2011intrinsic, cheng2017inhomogeneity, wu2017vortex, allmaras2018intrinsic, vodolazov2019minimal} and experimentally \cite{you2013jitter, calandri2016superconducting, wu2017improving, caloz2018high, sidorova2018intrinsic, korzh2018wsi}. Many efforts have been implemented in order to minimize extra jitter due to electrical noise. Since the main focus has been on the local jitter inherent in the detection process, experimental studies have been performed on short nanowires in order to eliminate the contribution of the nanowire length to the measured system jitter. Due to large kinetic inductance, long nanowire in a typical SNSPD behaves as a high-impedance transmission line with a phase velocity of only 0.03$c$ ($c$ the speed of light in vacuum) \cite{santavicca2016microwave, zhao2018distributed}. Therefore, in practical SNSPD devices, contributions due to detector geometry (nanowire length) and due to electrical noise often exceed together the local jitter. 
	\\\indent In our present study we evaluate the geometrical contribution to the system jitter in a straight nanowire with the topology similar to the one reported in Ref. \cite{sidorova2018timing}.  Furthermore, we show that large photon flux reduces drastically the geometrical contribution to the system jitter. The effect allows for calibration of the noise component of the jitter. Geometrical contribution to the system jitter was studied in a straight NbN nanowire by means of two methods: the differential technique and the first-order self-resonance. With these two methods we estimated propagation velocity of the electrical signal along the nanowire and found good agreement between our data and results of other groups that justifies self-consistency of our approach.
\section{Methods and Samples}
Differential or dual method for SNSPD readout was introduced in Ref. \cite{calandri2016superconducting, hofherr2014real}. It utilizes two identical readout channels for voltage pulses produced by the very same detection event. Pulses emerge from opposite ends of a current-carrying nanowire. They have opposite polarities and are recorded independently at two channels (Fig.~\ref{fig:sketch}a). 
\subsection{Formalism}
Following our previous concept \cite{sidorova2018timing}, we consider a photon count as a composite event including several sequential stages. Each stage either distorts the rising edge of the voltage pulse (noise) or adds its own random delay time to the experimentally measured relative arrival time of the voltage pulse at the recording instruments. The probability distribution of arrival times is described by the joint PDF. Hereafter STD of the joint PDF is used to characterize system jitter in order to circumvent the dependence of FWHM on the shape of PDF. Assuming no internal reflections and that $t_{1}$ and $t_{2}$ are random arrival times of pulses emerging at opposite edges of the nanowire and that they are measured at channels 1 and 2 (Fig.\ref{fig:sketch}a) with respect to the common laser trigger, we can express these times as sums:
\begin{figure}[t]
\center
\includegraphics[width=0.45\textwidth]{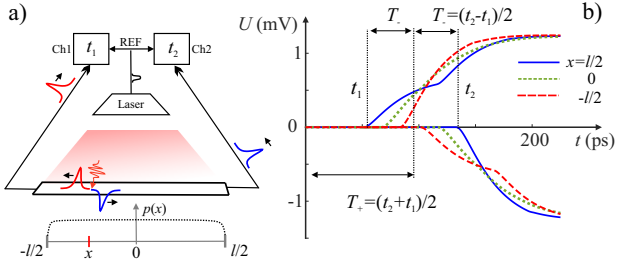}
\caption{\label{fig:sketch} Schematic representation of the pulse propagation in a straight nanowire (a). Simulated signal transients for a straight nanowire with hot spot parameters from Ref. \cite{zhao2018distributed, berggren2018superconducting} and three absorption positions $x = -l/2, 0$ and $l/2$ (b).}	
\end{figure}
%
\begin{equation}\label{eq:t-seq}
\left\{
\begin{aligned}
	t_{1} &= C_{1}+\tau_{\rm ins}+\tau_{\rm opt}+\tau_{\rm loc}+x/\nu_{\rm p}+\tau_{\rm n1}+\tau_{\rm ampl1} \\
	t_{2} &= C_{2}+\tau_{\rm ins}+\tau_{\rm opt}+\tau_{\rm loc}+(l-x)/\nu_{\rm p}+\tau_{\rm n1}+\tau_{\rm ampl1} 	
\end{aligned}
\right.
\end{equation}
where fixed constants $C_{1}$ and $C_{2}$ are average values of $t_{1}$ and $t_{2}$, which are defined by the lengths of the optical and of the readout paths. 
The delay term $\tau_{\rm ins}$ in \eqref{eq:t-seq} describes instrumental jitter. It represents random lags between optical pulse of the laser, laser trigger and the readout clock. The term $\tau_{\rm opt}$ stays for the optical jitter that arises due to random traveling time of a photon between the laser and the absorption site in the nanowire \cite{sidorova2018intrinsic}. The term $\tau_{\rm loc}$ represents random delay time (latency) inherent in the detection mechanism. This delay time produces local jitter and is equal for both readout channels.

We assign each photon absorption site in the nanowire of the length $l$ the longitudinal random variable $x\ni[-l/2;l/2]$ (Fig. 1a) which obeys uniform probability distribution as long as the probability of photon absorption is the same at all absorption sites. For voltage pulses arriving at channels 1 and 2, this distribution randomizes their traveling times along the nanowire: $x/\nu_{\rm p}$ and $(l-x)/\nu_{\rm p}$. Here $\nu_{\rm p}$ is the propagation speed of electrical signals along the nanowire. Times $\tau_{\rm n1}$ and $\tau_{\rm n2}$ describe distortion of rising edges for pulses at channels 1 and 2, respectively, due to electrical noise in these channels. The noise contributions in two channels are not correlated. Times $\tau_{\rm ampl1}$  and  $\tau_{\rm ampl2}$ are random delays, due to fluctuation of pulse amplitudes (i.e. time-walk; these delays should be correlated, since if both channels have the same bandwidth and gain). 

Arrival times $t_{1}$ and $t_{2}$ can be expressed through symmetric and anti-symmetric components as $t_{1}=T_{+}+T_{-}$ and $t_{2}=T_{+}-T_{-}$, where  the symmetric component  $T_{+}=(t_{1}+t_{2})/2$ is the average arrival time, and the anti-symmetric component $T_{-}=(t_{1}-t_{2})/2$  is the half-delay between two pulses (Fig.\ref{fig:sketch}b). Thus, we can write:
%
\begin{equation}\label{eq:T-seq}
\left\{
\begin{aligned}[l]
	T_{+} & = \left< T_{+} \right> + \tau_{\rm n+}  + \tau_{\rm ampl+} 
	 + \tau_{\rm ins} + \tau_{\rm opt} + \tau_{\rm loc} \\
	T_{-} &= \left< T_{-} \right> + \tau_{\rm n-} +\tau_{\rm ampl-} + x/\nu_{\rm p}
\end{aligned}
\right.
\end{equation}
where $ \left< T_{+} \right> = (C_{1}+C_{2}+l/\nu_{\rm p})/2 $ and $ \left< T_{-} \right> = (C_{1}-C_{2}-l/\nu_{\rm p})/2 $ are constant average values, $\tau_{\rm n\pm} = (\tau_{\rm n1}\pm\tau_{\rm n2})/2$ and $\tau_{\rm ampl\pm} = (\tau_{\rm ampl1}\pm\tau_{\rm ampl2})/2$ are delays induced by noise and amplitude fluctuations correspondingly
The contribution to the jitter from amplitude fluctuations can be eliminated by defining the arrival times at a fixed fractional level on the rising edge of each voltage pulse (analogous to constant fraction discrimination). 
We evaluate STDs for $T_{+}$ and $T_{-}$ by building joint, compound PDFs for sequential independent random variables analogous to Eq. (B1) in Ref. \cite{sidorova2018timing}. Corresponding STDs are
%
\begin{equation}\label{eq:sigmas}
\left\{
\begin{aligned}
	\sigma_{T+} &= \sqrt{0.25\left( \sigma_{\rm n1}^2 + \sigma_{\rm n2}^2 \right) + \sigma_{\rm ins}^2 + \sigma_{\rm opt}^2 + \sigma_{\rm loc}^2}\\
	\sigma_{T-} &= \sqrt{0.25\left( \sigma_{\rm n1}^2 + \sigma_{\rm n2}^2 \right) + \sigma_{\rm geom}^2}
\end{aligned}
\right.
\end{equation}
where $ \sigma_{\rm geom} = \sigma_{x}/\nu_{\rm p} $ is the geometrical jitter caused by position dependent traveling times of voltage pulses from the absorption site. 
Measuring   and defining noise,   in two channels independently, one can obtain geometrical jitter. According to Ref. \cite{santavicca2016microwave, zhao2017single, zhao2018distributed} the propagation speed $ \nu_{\rm p} $  of the voltage pulse in a nanowire is $\nu_{\rm p} \approx (0.03 \pm 0.01)c $. Thus, the geometrical jitter should linearly depend on the length $l$ of the nanowire. For uniformly distributed probability of photon absorption and in the absence of noise, PDF of the geometrical jitter is also uniform with the FWHM $\Delta t \approx l/\nu_{\rm p}$ and STD $\sigma = \Delta t / 2\sqrt3 $ . For a nanowire length $l = 180$~$\mu m$ this results in a geometrical jitter  $6\pm2$~ps. The above consideration neglects internal reflections of the electrical pulses at the boundaries of the hot spot and at the edges of the nanowire. As it was shown in Ref.\cite{zhao2018distributed} , these reflections make apparent jitter dependent on the threshold level, which is used to define arrival times of pulses at channels 1 and 2. Fig. \ref{fig:sketch}b shows the effect of reflections on the rising edges of simulated signal transients at the inputs of both channels.
\subsection{Contribution of electrical noise}
For a fixed threshold trigger level on the rising edge of the pulse with a slew rate SR = $dV/dt$, presence of an instant positive noise voltage causes earlier trigger than it would happen in the absence of the noise. Correspondingly, the apparent pulse arrival time will be less than without noise. The rms difference between these two arrival times is called noise jitter \cite{sidorova2017physical}. If the rms voltage noise at the input of the readout channel is given by $\sigma_{U \rm n} $, the noise jitter can be computed as:  
%
\begin{equation}\label{eq:noisejitter}
 \sigma_{\rm n} = \sigma_{U \rm n}/ \rm SR
\end{equation}
The mean slew rate could be estimated as SR $\approx 0.8A/\tau_{\rm rise} $ where $A$ is the mean amplitude of the pulse and $\tau_{\rm rise}$ is the mean rise time of the pulse measured between 0.1$A$ and 0.9$A$. 
Apart from jitter measurements, the noise jitter $ \sigma_{\rm n}$  can be estimated from the expected slew rate of detector pulses along with the effective noise temperature and the bandwidth of the readout as $  \sigma_{\rm n} = \sqrt{ k_{\rm B} T_{\rm n} Z_{0} B G}/\rm SR $ , where $k_{\rm B} =1.38 \times 10^{-23}$~J/K is the Boltzmann  constant, $T_{\rm n}$  is an effective noise temperature of the readout seen by the nanowire, $Z_{0}$ = 50~$\Omega $ is the characteristic impedance of the readout, $B$ is the effective bandwidth of the noise and $G$ is the total power gain of the readout channel. To estimate relative contributions to the noise temperature of the whole readout, one could use Friis formula: $T_{\rm n} = T_{1} + T_{2}/G_{1} + T_{3}/G_{1}G_{2}+ \dots $  , where $T_i$ and $G_i$ are noise temperatures and gains of elements in the readout chain. A significant contribution to $T_{\rm n}$ could arise due to discretization noise of the analog-to-digital converter (ADC). This contribution $T_{\rm ADC}$ can be estimated with the known signal-to-noise ratio (SNR) of the converter as  $ \rm SNR =$ $\left( V_{\rm FS} / \sigma_{U \rm n} \right)^2 \approx 2^{2n}$ , where $V_{\rm FS}$ is the voltage of the full scale, and $n$ is the resolution in bits. 

If the intrinsic bandwidth of the pulses produced by the nanowire is smaller than the readout bandwidth, $0.35/\tau_{\rm int} \ll B$, the slew rate is independent of the readout and the noise jitter reduces with the reduction of $B$ as $\sigma_{\rm n} \propto \sqrt{B} $. In the opposite case, when the rise time of the voltage pulse on the oscilloscope is determined solely by the readout bandwidth, $\tau_{\rm rise} \equiv 0.35/B > \tau_{\rm int} $, the noise jitter increases with the reduction of the bandwidth as $\sigma_{\rm n} \propto 1/ \sqrt{B} $ . Thus, it is important to choose an optimal bandwidth to minimize noise jitter.
\subsection{Straight 180-$\mu$m-long NbN nanowires}
We have fabricated our straight nanowires from a NbN film with a thickness of $d = 5$~nm on a sapphire substrate. The film was deposited using reactive magnetron sputtering onto the heated substrate. For pattering of the nanowire, we employed a negative PMMA technique \cite{charaev2017enhancement} . The nanowires have a width of $w \approx 105$~nm and a length $l = 180 $~$\mu$m. The critical temperature $T_{\rm c} = 11.2$~K and the critical current $I_{\rm c} = 28.5 - 33 $~$\mu$A were measured. The design of the samples is shown in Fig. 2. The nanowire is connected at both sides to a coplanar waveguide with a characteristic impedance of $Z_0 = 50$~$\Omega$ (Fig. \ref{fig:exp} a). The nanowire has tapers to prevent current crowding in contact areas (Fig.\ref{fig:exp} b). In order to prevent optical diffraction, we additionally patterned an array of disconnected nanowires (Fig. \ref{fig:exp} c) parallel to the photon detecting nanowire.
\section{Measurements}
\begin{figure}[t]
\center
\includegraphics[width=0.4\textwidth]{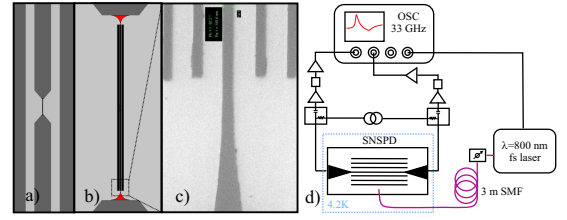}
\caption{\label{fig:exp} Design of the sample for the differential readout (a). Design of the central part with a straight SNSPD (b). Scanning-electron microscopy image of the nanowires near taper (c).  Schematic of the experimental setup (d).}	
\end{figure}

To ensure uniform illumination, the chip was mounted in a copper detector block with two SMA connectors and a fiber output located >5 mm apart from the substrate with the nanowire. The end of the fiber was mounted in the adjustable Teflon sled, which allows for fine lateral positioning of the fiber against the detector at the operation temperature right before measurements. To deliver optical pulses to the nanowire we used a single-mode fiber. Schematic of the setup is shown in Fig. \ref{fig:exp} d. The detector block was connected via two coaxial cables to two room-temperature broadband bias-tees (Antritsu K251) with low input losses. The dc ports of bias-tees were connected to a battery-driven low-noise current source for detector bias. To amplify voltage pulses output by the nanowire, we used for both readout channels identical low-noise MITEQ AFS amplifiers. 
The amplifiers have a noise temperature $T_{\rm n1} \approx 80$~K, bandwidth $B_1 = 8$~GHz and a gain $G_1 = 40$~dB. For the second stage we used MITEQ amplifiers of AMF series with 6-dB input attenuators, $T_{\rm n1} \approx 1200$ K and $G_2 = 24$~dB with a maximum peak output voltage of 1 V. Total voltage amplification is about 1550. As the recording instrument we used the real-time oscilloscope Keysight Infiniium X93204A with a bandwidth $B_3 = 5$ GHz (10 GSa/s at 12 bit) and $\approx $ 1~mV rms noise at 100 mV/division ($T_{\rm n1} \approx 2\times 10^5$~K). In this setup the noise from the first-stage amplifier should dominate. With the effective bandwidth of the setup $B_{\rm eff} \approx 5$ GHz the effective noise temperature $T_{\rm n} = 125 \pm 10$~K is expected which should result in an rms noise voltage  $ \sigma_{U \rm n} \approx 32$~mV.
%
\subsection{Triggering, noise jitter, $T_-$ and $T_+$ jitter}
The real-time oscilloscope continuously digitizes signals at the input channels and stores all the measured points in the internal cyclic buffer with time stamps which are positioned with an accuracy better than 100~fs. Triggering was performed at the rising edge of the voltage pulse from one of the nanowire ends. The trigger was not used for measurements of the interval between arrival times of pulses, but only to select a relevant data from the cyclic buffer of the oscilloscope.

\begin{figure}[t]
\center
\includegraphics[width=0.47\textwidth]{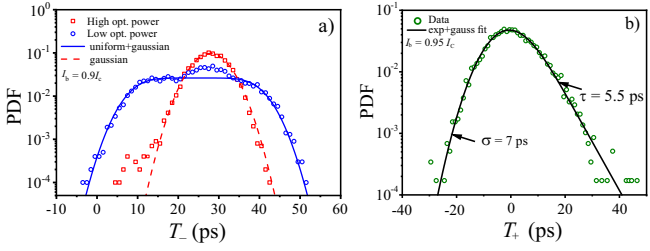}
\caption{\label{fig:exp_jitt} Experimental PDFs for the time difference $T_{-}$; open circles - low optical power, open rectangles - high optical power. Solid lines are fits. (a) Typical histogram for the time delay $T_{+}$ at the small light intensity. Solid curve is a fit with an exponentially modified Gaussian distribution (b).}	
\end{figure}

To eliminate extra jitter due to amplitude fluctuations, we measured all arrival times at the fixed fractional level on the rising edge of the corresponding voltage pulse. To measure time intervals, we used a built-in function of the oscilloscope for delta-time measurement. This function finds the selected level of each instantaneously acquired pulse and measures the time interval between pulses at different channels. Single measurement gives instantaneous values of $t_1$ and $t_2$ with respect to the laser trigger, which are stored for further processing. We accumulated $10^4$ measurements and obtained histograms shown in Fig. \ref{fig:exp_jitt} .

Estimated optical jitter due to the dispersion in the fiber is $\sigma_{\rm opt} \approx 8 $~ps \cite{sidorova2018timing}. The measured instrumental jitter is    $\sigma_{\rm ins} \approx 1.8 $~ps. The measured voltage noise is $ \sigma_{U \rm n1} \approx 26$~mV  and $ \sigma_{U \rm n2} \approx 40$~mV   for channel 1 and 2, correspondingly, which is in good agreement with the estimations. The slew rates at rising edges were $ \rm SR_1 \approx 6$~mV/ps and $ \rm SR_2 \approx 6.5$~mV/ps, which results in noise jitters of $ \sigma_{\rm n1} \approx 4.0 \pm 0.8$~ps and $ \sigma_{\rm n1} \approx 6$~ps . The rise time of pulses is $\tau_{\rm rise} \approx $ 125~ps, which is approximately corresponds to the expected time resolution of the readout. The $T_-$ histograms shown in Fig. \ref{fig:exp_jitt}a were recorded at two different incident photon fluxes on the nanowire.
The histograms have notably different shapes. An optical attenuator, which we used in the experiment, was not calibrated that leaves the relative light intensity unknown. At the larger intensity we obtained an almost Gaussian PDF with the standard deviation $\sigma_{T-} \approx 4$~ps. At the lower intensity STD was significantly larger  $\sigma_{T-} \approx 9.4$~ps, while PDF has a non-gaussian, flat profile.

Fig. \ref{fig:exp_jitt}b shows a typical $T_+$ histogram recorded at the lower light intensity and the best fit with the exponentially modified Gaussian distribution described in Ref. \cite{sidorova2018timing} . 
%
\subsection{First order self-resonance}
\begin{figure}[t]
\center
\includegraphics[width=0.26\textwidth]{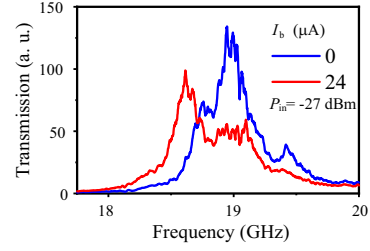}
\caption{\label{fig:resonance} Microwave transmission spectrum of the nanowire at two bias currents}	
\end{figure}

According to electromagnetic simulations with our detector parameters, nanowire has a characteristic impedance $Z_{\rm wire} \approx 3$ k$\Omega$ as a transmission line. We do not use matching circuits between the nanowire and the readout; the nanowire ends were directly coupled to 50-$\Omega $ lines. It allows to observe self-resonances in the nanowire, which is an inductively coupled distributed transmission resonator. According to the reported values of the propagation speed  \cite{santavicca2016microwave, zhao2018distributed}, the first-order self-resonance ($l = \lambda /2$,  is the wavelength) for our nanowire with the length $l$ = 180~$\mu$m is expected at the frequency $f_{\lambda/2} = \nu_{\rm p}/2l$ in the range $17 \pm 3$~GHz. To probe the self-resonance, we removed the amplifiers and connected the nanowire to a vector network analyzer. The resonance was found at a central frequency of  $f_{\lambda/2} = 18.5$~GHz (Fig. \ref{fig:resonance}). The resonance frequency decreases with the increase of the dc bias current that corresponds to an increase of the kinetic inductance.  
\section{Discussion}
The PDF of the half-delay time between arrivals of pulses at two channels, $T_-$, which is shown in Fig.~\ref{fig:exp_jitt}a with open circles, was obtained at the smaller light intensity. At small intensities the nanowire operates in the discrete single-photon detection regime. Recorded histogram contains both the geometrical and noise contributions (Eq. \eqref{eq:sigmas}). In the following discussion we neglect internal reflection in the nanowire. We believe that due to saturation of the 2nd amplifier and corresponding gain compression, measurements were performed effectively on a lower than 50\% threshold level. This partly compensated the effect of internal reflections (see Fig. 1b). Another important issue is the resistance and the life time of the normal domain in the nanowire which are also affected by reflections.
Suggesting that noise jitter and geometrical jitter are statistically independent, we estimated geometrical jitter from Eq. \eqref{eq:sigmas} as $\sigma_{\rm geom} = \sqrt{ \sigma_{T-}^2 - 0.25\left( \sigma_{\rm n1}^2  + \sigma_{\rm n2}^2\right) } \approx$ 8.5~ps . Standard deviation $\sigma_{T-}$ was numerically computed from the experimental histogram.

Further, assuming uniform probability of photon absorption and neglecting reflections in the nanowire, we estimate propagation speed as $\nu_{\rm p} = l/2\sqrt3 \sigma_{\rm geom} \approx 6.2\times10^6 $~m/s or 0.02$c$. To do it more rigorously, we fit the measured PDF with the convolution of the uniform and the Gaussian probability distributions, which describe two conditionally independent variables (absorption position and voltage noise):
%
\begin{equation}\label{eq:erf}
\text{PDF} = \frac{H}{2} \left[ \text{erf} \left( \frac{ T_{-} - t_0 }{ \sqrt{2} \sigma_{\rm n}} \right) - \text{erf} \left( \frac{ T_{-} - t_0 - l/ \nu_{\rm p} }{ \sqrt{2} \sigma_n} \right) \right]
\end{equation}
Here erf is the error function and $H$, $t_0$ are the height and the center of the histogram, $\sigma_{\rm n}$ is the measured noise jitter and $\nu_{\rm p}$  is the fitting parameter. The best fit is shown in Fig.~\ref{fig:exp_jitt}a (open circles). It was obtained with the propagation speed $ \nu_{\rm p} \approx 6 \times 10^6$~m/s. 
 The PDF of the half-delay time (Fig. \ref{fig:exp_jitt}a, open squares) obtained with the larger light intensity was fitted with the Gaussian probability distribution. The best fit shown in Fig. \ref{fig:exp_jitt}a has the STD of 4 ps, which coincides with the calculated value of the noise jitter $\sigma_{\rm n} \approx 0.5\sqrt{\sigma_{\rm n1}^2+\sigma_{\rm n2}^2}$. We associate the emergence of the Gaussian PDF with the transit of the nanowire into the bolometric operation regime as it was described in Ref. \cite{sidorova2018timing}. In this regime, the geometrical jitter is averaged out.
The propagation speed obtained with the fitting procedure agrees well with the value obtained from $\lambda/2$ self-resonance.
%
\section{Conclusion}
We have found an almost uniformly distributed difference between times of emergence of photon counts at two ends of a straight nanowire subject to uniform photon flux and directly coupled to 50-$\Omega$ environment. We attribute the uniform distribution to the geometrical jitter and obtained values of the propagation speed for voltage pulses in the nanowire in good agreement with the previously reported values and with the speed, which was estimated from the central frequency of a first-order self-resonance. We associate the reduction of the geometrical jitter below the noise jitter at large photon fluxes with the emergence of the bolometric detection regime, which offers the possibility for accurate calibration of the contribution of the electrical noise into the timing jitter.
\section*{Acknowledgment}
Authors would like to thank I. Charaev (MIT) and In-Ho Baek for the help with films preparation and measurements, A. Stassen and K.-H. Gutbrod for the help with preparation of samples and for fabrication of mechanical parts.
%
\bibliographystyle{IEEEtran}

\bibliography{Bibliography.bib}

\begin{thebibliography}{10}
\providecommand{\url}[1]{#1}
\csname url@samestyle\endcsname
\providecommand{\newblock}{\relax}
\providecommand{\bibinfo}[2]{#2}
\providecommand{\BIBentrySTDinterwordspacing}{\spaceskip=0pt\relax}
\providecommand{\BIBentryALTinterwordstretchfactor}{4}
\providecommand{\BIBentryALTinterwordspacing}{\spaceskip=\fontdimen2\font plus
\BIBentryALTinterwordstretchfactor\fontdimen3\font minus
  \fontdimen4\font\relax}
\providecommand{\BIBforeignlanguage}[2]{{%
\expandafter\ifx\csname l@#1\endcsname\relax
\typeout{** WARNING: IEEEtran.bst: No hyphenation pattern has been}%
\typeout{** loaded for the language `#1'. Using the pattern for}%
\typeout{** the default language instead.}%
\else
\language=\csname l@#1\endcsname
\fi
#2}}
\providecommand{\BIBdecl}{\relax}
\BIBdecl

\bibitem{sidorova2017physical}
M.~Sidorova, A.~Semenov, H.-W. H{\"u}bers, I.~Charaev, A.~Kuzmin, S.~Doerner,
  and M.~Siegel, ``Physical mechanisms of timing jitter in photon detection by
  current-carrying superconducting nanowires,'' \emph{Physical Review B},
  vol.~96, no.~18, p. 184504, 2017.

\bibitem{sidorova2018timing}
M.~Sidorova, A.~Semenov, H.-W. H{\"u}bers, A.~Kuzmin, S.~Doerner, K.~Ilin,
  M.~Siegel, I.~Charaev, and D.~Vodolazov, ``Timing jitter in photon detection
  by straight superconducting nanowires: Effect of magnetic field and photon
  flux,'' \emph{Physical Review B}, vol.~98, no.~13, p. 134504, 2018.

\bibitem{zhao2011intrinsic}
Q.~Zhao, L.~Zhang, T.~Jia, L.~Kang, W.~Xu, J.~Chen, and P.~Wu, ``Intrinsic
  timing jitter of superconducting nanowire single-photon detectors,''
  \emph{Applied Physics B}, vol. 104, no.~3, pp. 673--678, 2011.

\bibitem{cheng2017inhomogeneity}
Y.~Cheng, C.~Gu, and X.~Hu, ``Inhomogeneity-induced timing jitter of
  superconducting nanowire single-photon detectors,'' \emph{Applied Physics
  Letters}, vol. 111, no.~6, p. 062604, 2017.

\bibitem{wu2017vortex}
H.~Wu, C.~Gu, Y.~Cheng, and X.~Hu, ``Vortex-crossing-induced timing jitter of
  superconducting nanowire single-photon detectors,'' \emph{Applied Physics
  Letters}, vol. 111, no.~6, p. 062603, 2017.

\bibitem{allmaras2018intrinsic}
J.~P. Allmaras, A.~G. Kozorezov, B.~A. Korzh, and M.~D. Shaw, ``Intrinsic
  timing jitter and latency in superconducting single photon nanowire
  detectors,'' \emph{arXiv preprint arXiv:1805.00130}, 2018.

\bibitem{vodolazov2019minimal}
D.~Y. Vodolazov, ``Minimal timing jitter in superconducting nanowire
  single-photon petectors,'' \emph{Physical Review Applied}, vol.~11, no.~1, p.
  014016, 2019.

\bibitem{you2013jitter}
L.~You, X.~Yang, Y.~He, W.~Zhang, D.~Liu, W.~Zhang, L.~Zhang, L.~Zhang, X.~Liu,
  S.~Chen \emph{et~al.}, ``Jitter analysis of a superconducting nanowire single
  photon detector,'' \emph{Aip Advances}, vol.~3, no.~7, p. 072135, 2013.

\bibitem{calandri2016superconducting}
N.~Calandri, Q.-Y. Zhao, D.~Zhu, A.~Dane, and K.~K. Berggren, ``Superconducting
  nanowire detector jitter limited by detector geometry,'' \emph{Applied
  Physics Letters}, vol. 109, no.~15, p. 152601, 2016.

\bibitem{wu2017improving}
J.~Wu, L.~You, S.~Chen, H.~Li, Y.~He, C.~Lv, Z.~Wang, and X.~Xie, ``Improving
  the timing jitter of a superconducting nanowire single-photon detection
  system,'' \emph{Applied optics}, vol.~56, no.~8, pp. 2195--2200, 2017.

\bibitem{caloz2018high}
M.~Caloz, M.~Perrenoud, C.~Autebert, B.~Korzh, M.~Weiss, C.~Sch{\"o}nenberger,
  R.~J. Warburton, H.~Zbinden, and F.~Bussi{\`e}res, ``High-detection
  efficiency and low-timing jitter with amorphous superconducting nanowire
  single-photon detectors,'' \emph{Applied Physics Letters}, vol. 112, no.~6,
  p. 061103, 2018.

\bibitem{sidorova2018intrinsic}
M.~Sidorova, A.~Semenov, A.~Kuzmin, I.~Charaev, S.~Doerner, and M.~Siegel,
  ``Intrinsic jitter in photon detection by straight superconducting
  nanowires,'' \emph{IEEE Transactions on Applied Superconductivity}, vol.~28,
  no.~7, pp. 1--4, 2018.

\bibitem{korzh2018wsi}
B.~Korzh, Q.-Y. Zhao, S.~Frasca, D.~Zhu, E.~Ramirez, E.~Bersin, M.~Colangelo,
  A.~Dane, A.~Beyer, J.~Allmaras \emph{et~al.}, ``Wsi superconducting nanowire
  single photon detector with a temporal resolution below 5 ps,'' in
  \emph{CLEO: QELS\_Fundamental Science}.\hskip 1em plus 0.5em minus
  0.4em\relax Optical Society of America, 2018, pp. FW3F--3.

\bibitem{santavicca2016microwave}
D.~F. Santavicca, J.~K. Adams, L.~E. Grant, A.~N. McCaughan, and K.~K.
  Berggren, ``Microwave dynamics of high aspect ratio superconducting nanowires
  studied using self-resonance,'' \emph{Journal of Applied Physics}, vol. 119,
  no.~23, p. 234302, 2016.

\bibitem{zhao2018distributed}
Q.-Y. Zhao, D.~F. Santavicca, D.~Zhu, B.~Noble, and K.~K. Berggren, ``A
  distributed electrical model for superconducting nanowire single photon
  detectors,'' \emph{Applied Physics Letters}, vol. 113, no.~8, p. 082601,
  2018.

\bibitem{hofherr2014real}
M.~Hofherr, \emph{Real-time imaging systems for superconducting nanowire
  single-photon detector arrays}.\hskip 1em plus 0.5em minus 0.4em\relax KIT
  Scientific Publishing, 2014, vol.~16.

\bibitem{berggren2018superconducting}
K.~K. Berggren, Q.-Y. Zhao, N.~Abebe, M.~Chen, P.~Ravindran, A.~McCaughan, and
  J.~C. Bardin, ``A superconducting nanowire can be modeled by using spice,''
  \emph{Superconductor Science and Technology}, vol.~31, no.~5, p. 055010,
  2018.

\bibitem{zhao2017single}
Q.-Y. Zhao, D.~Zhu, N.~Calandri, A.~E. Dane, A.~N. McCaughan, F.~Bellei, H.-Z.
  Wang, D.~F. Santavicca, and K.~K. Berggren, ``Single-photon imager based on a
  superconducting nanowire delay line,'' \emph{Nature Photonics}, vol.~11,
  no.~4, p. 247, 2017.

\bibitem{charaev2017enhancement}
I.~Charaev, T.~Silbernagel, B.~Bachowsky, A.~Kuzmin, S.~Doerner, K.~Ilin,
  A.~Semenov, D.~Roditchev, D.~Y. Vodolazov, and M.~Siegel, ``Enhancement of
  superconductivity in nbn nanowires by negative electron-beam lithography with
  positive resist,'' \emph{Journal of Applied Physics}, vol. 122, no.~8, p.
  083901, 2017.

\end{thebibliography}

\end{document}